# Dirac Notation, Fock Space and Riemann Metric Tensor in Information Retrieval Models

Xing M. Wang
Sherman Visual Lab, Sunnyvale, CA, USA

*Abstract*

Using *Dirac Notation* as a powerful tool, we investigate the three classical *Information Retrieval* (IR) models and some their extensions. We show that almost all such models can be described by vectors in *Occupation Number Representations* (ONR) of *Fock spaces* with various specifications on, e.g., occupation number, inner product or term-term interactions. As important cases of study, Concept Fock Space (CFS) is introduced for Boolean model; the basic formulas for *Singular Value Decomposition* (SVD) of *Latent Semantic Indexing* (LSI) Model are manipulated in terms of Dirac notation.  And, based on SVD, a *Riemannian metric tensor* is introduced, which not only can be used to calculate the relevance of documents to a query, but also may be used to measure the closeness of documents in data clustering.

**Table of Contents**

## Introduction

Dirac Notation has been widely used for vectors in Hilbert spaces of QM (Quantum Mechanics) ([1], §7.2). Recently, it has also been introduced into Information Retrieval (IR) [19].

In this article, we apply Dirac notation to investigate the three classical IR models (vector space, Boolean and probabilistic) and some their extensions [2-3]. We find that most of them can be described by vectors in *Occupation Number Representations* (ONR) of Fock spaces, used in second quantization ([1], §22.1). In this description, each term of the data collection represents a possible state of a single "particle"; each document is a system of "particles", representing which "term-state" is occupied by how many "particles". If occupation number can only be 0 or 1, we have Boolean model in *Concept Fock Space* (CFS), like a system of fermions in QM. If occupation number can be any non-negative integer, we have the *Term Frequency Fock Space* used in Singular Value Decomposition (SVD) for Latent Semantic Indexing (LSI) Model [11-12], like a system of bosons in QM. Other models can be derived either by relaxing the requirement on the expectation value, or changing the definition of the weight, or introducing term-term interaction…

As a case study, we use Dirac notation to derive and express a set of basic formulas in SVD for LSI Model. A Riemannian metric tensor is introduced based on these formulas. We use it to calculate the relevance of documents to a query. We also propose to use the same metric to define the *neighborhoods* of documents in data space, by mapping them to a unit sphere in the metric space and measuring their distances. This map might be used in data clustering, like the diffusion maps in Ref. [20-21].

Whenever is possible, we will make use of or make comparison to Ref. [19]. To give numerical results, we often use the famous example of [3] (See Appendix A).

# 1. The Classical IR Models and Fock Spaces

In this section, we first give detailed applications of Dirac notation for the three classical models: vector space model, Boolean model and probabilistic model in their classical ways. Then we introduce Fock space as a unified way to represent the above models.

## 1.1. The Vector Space Model and Weighted Term Vector Space

Suppose we have *t-independent terms* in our collection, we can think of them as othonormal vectors in a t-dimensional real vector space:

$$\vec{k}_i: \qquad \vec{k}_i \cdot \vec{k}_j = \delta_{ij} = \begin{cases} 0, if\ i \neq j \\ 1, if\ i = j \end{cases}, \quad i, j \in \{0,1,...t\} \tag{1.1.1a}$$

In *Dirac notation*, we can define a set of the orthonormal vectors $|k_i\rangle$ ([1], pages 192-195):

$$|k_i\rangle: \quad \langle k_i| = (|k_i\rangle)^T \ ; \quad \langle k_i|k_j\rangle = \delta_{ij} \tag{1.1.1b}$$

Here, $|k_i\rangle$ is called a *ket*; $\langle k_i|$ is its Hermit conjugate (or, transpose for real vector), and is called a *bra*; $\langle k_i|k_j\rangle$ is a bracket, the *inner product of two vectors*. More about Dirac notation can be found in Ref [19] and [20].

Because any vector in the term-document space (i.e., terms, documents and queries) can be uniquely expanded with respect to the terms, we say that the terms construct a basis of the space. Moreover, if the expansion coefficients are uniquely determined by the inner product of document (and query) with term, the completeness can be written as:

$$\sum_{i=1}^{t} |k_i\rangle\langle k_i| = I \tag{1.1.1c}$$

Now we can represent a query vector $q$ and document $d_\mu$ (μ=1, 2…N) as:

$$|q\rangle = I|q\rangle = \sum_{i=1}^{t}|k_i\rangle\langle k_i|q = \sum_{i=1}^{t} w_{i,q}|k_i\rangle \ ; \quad |d_\mu\rangle = \sum_{i=1}^{t} w_{i,\mu}|k_i\rangle. \tag{1.1.2a}$$

Here, $w_{i,q}$ and $w_{i,\mu}$ are called *term weights* of a query or a document with respect to the i[th] term. They *actually define* the inner product of a vector (or a query) with a term ([2], page 29; [3], page 15):

$$\langle k_i|q\rangle = \langle q|k_i\rangle = w_{i,q} \ ; \langle k_i|d_\mu\rangle = \langle d_\mu|k_i\rangle = w_{i,\mu} \tag{1.1.2b}$$

Note that we have used the fact that, for two real vectors,

$$\langle u | v \rangle = \langle v | u \rangle$$

This is the *classical vector space model (VSM)*, proposed by Salton et al ([4, 5]). We call the vector space spanned by these t base-vectors of Eq. (1.1.1b) a *Weighted Term Vector Space*, or WT-space for short. The weights can be calculated based on different schemes, usually involving ([2], page 29; [3], page 15) following parameters:

$tf_{i,\mu}$ : The frequency of *i-th* term in document $\mu$         (1.1.3a)

$n_i$ : The number of documents containing *i-th* term         (1.1.3b)

$dl_\mu$ : The length (or word count) of the $\mu$-th document         (1.1.3c)

$N$ : Number of all documents in the collection         (1.1.3d)

$idf_i$ : The so-called *inverse document frequency* (idf) of *i-th* term         (1.1.3e)

$tf_{i,q}$ : The frequency of *i-th* term in the query ([2], page 29)         (1.1.3f)

In reference [3], the following formula is used to calculate *idf*, $w_{i,\mu}$ and $w_{i,q}$:

$$idf_i = \log\left(\frac{N}{n_i}\right) \tag{1.1.4a}$$

$$w_{i,\mu} = tf_{i,\mu} \cdot idf_i \tag{1.1.4b}$$

$$w_{i,q} = tf_{i,q} \cdot idf_i \tag{1.1.4c}$$

Many variations on how to compute weights have been done. One good performer is identified as ([3], page 17):

$$w_{i,\beta} = \frac{(\log tf_{i,\beta} + 1.0) \cdot itf_i}{\sqrt{\sum_{j=1}^{t}\left[(\log tf_{i,\beta} + 1.0) \cdot itf_i\right]^2}} \tag{1.1.4d}$$

After we have computed the weights, we now can calculate the relevance or the Similarity Coefficient (SC) of document $d_\mu$ with respect to query $q$. In classical vector model, SC is the cosine of the angle between the two vectors, ([2], page 27), which can be written as:

$$SC(d_\mu, q) = \frac{\langle d_\mu | q \rangle}{|d_\mu| \cdot |q|} = \frac{\langle d_\mu | q \rangle}{\sqrt{\langle d_\mu | d_\mu \rangle} \cdot \sqrt{\langle q | q \rangle}} \tag{1.1.5}$$

Here, using (1.1.2a, b) and (1.1.1c), we have:

$$\langle d_\mu | q \rangle = \langle q | d_\mu \rangle = \langle d_\mu | I | q \rangle = \sum_{i=1}^{t} \langle d_\mu | k_i \rangle \langle k_i | q \rangle$$
$$= \sum_{i,k=1}^{t} w_{i,\mu} \langle k_i | k_k \rangle w_{k,q} = \sum_{i,k=1}^{t} w_{i,\mu} \delta_{i,k} w_{k,q} = \sum_{i=1}^{t} w_{i,\mu} w_{i,q}. \tag{1.1.6}$$

$$|d_\mu|^2 = \langle d_\mu | d_\mu \rangle = \sum_{i,k=1}^{t} w_{i,\mu} \langle k_i | k_k \rangle w_{k,j} = \sum_{i,k=1}^{t} w_{i,\mu} \delta_{i,k} w_{k,j} = \sum_{i=1}^{t} w_{i,\mu}^{\,2}. \tag{1.1.7a}$$

$$|q|^2 = \langle q | q \rangle = \sum_{i,k=1}^{t} w_{i,q} \langle k_i | k_k \rangle w_{k,q} = \sum_{i,k=1}^{t} w_{i,q} \delta_{i,k} w_{k,q} = \sum_{i=1}^{t} w_{i,q}^{\,2}. \tag{1.1.7b}$$

In many cases, we want to normalize $d_j$ and $q$, so that:

$$|d_\mu|^2 = \langle d_\mu | d_\nu \rangle = \sum_{i=1}^{t} w_{i,\nu}^2 = 1; \quad |q|^2 = \langle q | q \rangle = \sum_{i=1}^{t} w_{i,q}^2 = 1 \tag{1.1.8}$$

We see that the *normalized WT-space* is a *bounded t-dimensional continuous space* over the field of [0, 1], restricted in the unit cubes:

$$|w\rangle = \sum_{i=1}^{t} w_i | k_i \rangle, \quad w_i \in [0,1] \subset \Re \tag{1.1.9}$$

Note that, the vectors restricted by Eq. (1.1.9) *do not form a vector space*, because their inverse elements are not included, and they do not close to vector addition. But they do form a subspace, closed to retrieval-related operations. We call such space *a retrieval-closed space*. Actually, this space is better to be classified as a Fock space (see Section 1.3).

Up to now, the only properties (or assumptions) we know about the base vectors $|k_i\rangle$ are in Eq. (1.1.1). These assumptions may be relaxed or violated in other models. We also have to define the rules to calculate the inner product of document and query with term vector (the weights), for example, by Eq. (1.1.4).

If we following the discussion in Ref [19], we can rewrite Eq. (1.1.6) using the trace operation:

$$\langle d_\mu | q \rangle = tr(| d_\mu \rangle\langle q |) = \sum_{i=1}^{t} \langle k_i | d_\mu \rangle\langle q | k_i \rangle \tag{1.1.10}$$

We accept it as Relevance Coefficient (*the cosine law*), while in Ref [19], the related probability is given by (page 87) the *cosine-squared law*:

$$tr(| q \rangle\langle q | d_\mu \rangle\langle d_\mu |) = |\langle q | d_\mu \rangle|^2 \tag{1.1.11}$$

In reference [3], a simple example is used throughout that book. To make it convenient for readers, we will reuse this example and restate the example ([3], page 15). The example, referred to *GF-Example* from now on, has a collection of three Documents and one Query as described in *Appendix A*.

To avoid unnecessary confusion, we use the following conventions for labeling indexes of vectors or tensors:

**Index Convention**:
1. For terms, we use Latin chars $i, j, k, l, m$…
2. For documents, we use Greek chars $\alpha, \beta, \mu, \nu$…
3. For query, we use Latin char $q$.
4. For other cases, like in LSI, we use Latin chars $a, b, c$…

**1.2. The Classical Boolean Model and Boolean Term Vector Space**

The classical Boolean model is discussed in [2], §2.5.2. Here we want to use a vector space to describe this model.

Assume that the term-document and term-query weights are all binary, i.e.:

$$w_{i,\mu} = \langle k_i | d_\mu \rangle \in \{0, 1\} \; ; \quad w_{i,q} = \langle k_i | q \rangle \in \{0, 1\} \tag{1.2.1}$$

Now we introduce new vector space to represent documents and query in this space. Each vector is a director product of t 2-dimentional vector. This space has $M = 2^t$ points. We call this vector space a **Boolean Term Vector Space,** or **BT-space** for short. Suppose we have 5 independent terms, and a sample vector has three terms: first, second and the fifth. Then in the BT-space, this vector can be represented as:

$$
\begin{aligned}
| d \rangle &\equiv \prod_{\alpha=1}^{5} | n_\alpha \rangle_\alpha = |1\rangle_1 |1\rangle_2 |0\rangle_3 |0\rangle_4 |1\rangle_5 \\
&\equiv \begin{pmatrix}1\\0\end{pmatrix}_1 \begin{pmatrix}1\\0\end{pmatrix}_2 \begin{pmatrix}0\\1\end{pmatrix}_3 \begin{pmatrix}0\\1\end{pmatrix}_4 \begin{pmatrix}1\\0\end{pmatrix}_5 \equiv |1,1,0,0,1\rangle \; ;
\end{aligned} \tag{1.2.2a}
$$

Here we have used an explicit representation of each 2D factor space. They are chosen as eigenvectors of an *occupation operator*:

$$\hat{n}|1\rangle = |1\rangle, \quad \hat{n}|0\rangle = 0, \quad |1\rangle = \begin{pmatrix} 1 \\ 0 \end{pmatrix}, |0\rangle = \begin{pmatrix} 0 \\ 1 \end{pmatrix}, \quad \hat{n} = \begin{pmatrix} 1 & 0 \\ 0 & 0 \end{pmatrix} \qquad (1.2.2b)$$

The 2-dimentional vector looks similar to the spin-1/2 vector representation in quantum mechanics (QM). Only difference is, in QM, the base vectors usually are common eigenvectors of operators $\hat{s}^2$ and $\hat{s}_z$, corresponding to spin-up and spin-down states (in natural units, $\hbar = c = 1$):

$$\begin{aligned}\hat{s}_z|\uparrow\rangle &= \frac{1}{2}|\uparrow\rangle, & \hat{s}^2|\uparrow\rangle &= \frac{3}{4}|\uparrow\rangle \\ \hat{s}_z|\downarrow\rangle &= -\frac{1}{2}|\downarrow\rangle, & \hat{s}^2|\downarrow\rangle &= \frac{3}{4}|\downarrow\rangle\end{aligned} \qquad (1.2.3a)$$

Using these two states as basis, we have their matrix expressions:

$$|\uparrow\rangle = \begin{pmatrix} 1 \\ 0 \end{pmatrix}, \quad |\downarrow\rangle = \begin{pmatrix} 0 \\ 1 \end{pmatrix}, \quad \hat{s}_z = \frac{1}{2}\begin{pmatrix} 1 & 0 \\ 0 & -1 \end{pmatrix}, \quad \hat{s}^2 = \frac{3}{4}\begin{pmatrix} 1 & 0 \\ 0 & 1 \end{pmatrix} \qquad (1.2.3b)$$

Comparing (1.2.2b) and (1.2.3b), we see the following relations:

$$\hat{n} \sim \hat{s}_z + \frac{1}{2}, \quad |1\rangle \sim |\uparrow\rangle, \quad |0\rangle \sim |\downarrow\rangle \qquad (1.2.3c)$$

In BT-space, the basis in equation (1.1) now can be represented as a set of t elements of the form:

$$|k_i\rangle = \prod_{\alpha=1}^{t}|\delta_{i,\alpha}\rangle_\alpha = |0\rangle_1|0\rangle_2 \cdots |1\rangle_i \cdots |0\rangle_t = |0,0,\ldots,1,\ldots,0\rangle; \qquad (1.2.4a)$$

And we have t occupation operators:

$$\hat{n}_l|k_i\rangle = \delta_{li}|k_i\rangle, \quad l,i \in \{1,2,\ldots t\} \qquad (1.2.4b)$$

This means that $\hat{n}_i$ acts on the *l*-th 2-dimentional vector of the product; if it is |1>, returns an eigenvalue 1, otherwise returns 0. Therefore we can think of the t base vectors as the common eigenvectors of the t occupation operators. We call it occupation operator, because its eigenvalues are similar to the occupation number of a system of fermions (see §1.3).

A document now can be represented as:

$$|d_\mu\rangle = \prod_{i=1}^{t} |n_{i,\mu}\rangle_i \,;\; n_{i,\mu} = \langle k_i | d_\mu \rangle = w_{i,\mu} \in \{0, 1\}$$
$$\hat{n}_i | d_\mu \rangle = n_{i,\mu} | d_\mu \rangle$$
(1.2.5)

The documents in *GF-Example* now can be represented as:

$$|d_1\rangle = \prod_{i=1}^{t} |n_{1,i}\rangle_i = |1,0,1,0,1,1,1,1,1,0,0\rangle$$
$$|d_2\rangle = \prod_{i=1}^{t} |n_{2,i}\rangle_i = |1,1,0,1,0,0,1,1,0,1,1\rangle \quad (1.2.6a)$$
$$|d_3\rangle = \prod_{i=1}^{t} |n_{3,i}\rangle_i = |1,1,0,0,0,1,1,1,1,0,1\rangle$$

Now let us assume that our binary query is "$(Gold \wedge Shipment) \wedge \neg fire$", i.e.:

$$q = k_6 \wedge k_9 \wedge \neg k_5 \quad (1.2.6b)$$

Then we can see only $d_3$ is relevant, because it has the sixth and ninth term but not has the fifth term as the query requires. In other words, the relevant documents of this query are in the following set (or *a concept*):

$$D_q = \bigcup \{|x_1, x_2, x_3, x_4, 0, 1, x_7, x_8, 1, x_{10}, x_{11}\rangle, x_i \in \{0,1\}\} \quad (1.2.6c)$$

Using the vector space, we can also think that any relevant document ⟨d| is a common eigenvector of occupation operators $\hat{n}_5$, $\hat{n}_6$ and $\hat{n}_9$ such that:

$$\hat{n}_5 | d \rangle = | d \rangle, \hat{n}_6 | d \rangle = | d \rangle, \hat{n}_9 | d \rangle = 0 \quad (1.2.6d)$$

Our BT-space is very close to what is used by the *Generalized Vector Space*, introduced by Wang, Ziarko, and Wong [6]. The main difference is that they consider the space is of $2^t$ dimensional, and introduce $2^t$ base vector to represent a term vector and term-term correlations. In our case, we rather think it is a product of *t* factor spaces, which is spanned by the eigenvectors of the occupation operator g. This factor space is 2-dimesional for Boolean model, but can be extended easily, as described in next sections (see §1.3 and §2.1).

We see that the BT-space is a bounded, finite and *t-dimensional discrete space* over the field of {0, 1}, restricted to the vertices of the unit cubes in our WT-space:

$$|w\rangle = \sum_{i=1}^{t} w_i |k_i\rangle, \quad w_i \in \{0,1\} \subset \aleph \tag{1.2.7}$$

Again note that, the vectors defined by Eq. (1.2.7) do not form a vector space, but a retrieval-closed space of vectors. It is also a Fock space, as we will see in next section.

Following the discussion in [19] (pages 56-60), we see that our occupation operator is a projector, and can be written as:

$$\hat{n}_l = |k_l\rangle\langle k_l|, \quad \hat{n}_l |k_i\rangle = \delta_{li} |k_i\rangle, \quad l,i \in \{1,2,\ldots t\} \tag{1.2.8}$$

### 1.3. Fock Spaces and Occupation Number Representations

The vectors like (1.2.6a) remind us of the *Occupation Number Representation* (ONR) ([1], page 566) in a multi-particle system:

$$|\vec{N}\rangle = |N_1, N_1, \ldots N_t\rangle \tag{1.3.1}$$

Here $N_i$ is the number of particles occupying the $i^{th}$ state (orbital). This vector space is called *Fock space*. For fermions (e.g., electrons, protons…), $N_i$ can only be 0 or 1. Our BT-space is also a Fock space, and its representation can be called **Boolean Term Occupation Number Representation** (BT-ONR), in which the number can have only value 0 or 1 for a given term. The occupation operator $\hat{n}_i$ now can be interpreted as Term Occupation Number Operator. Its eigenvalue $N_i$ is 0 or 1 in BT-space. Because any term, word group, sentence, paragraph, document or query is represented by one of the M = $2^t$ vectors of the form (1.3.1), we can also think of TB-space as a finite, discrete set of $2^t$ points (or elements).

Because each point is a *concept*, therefore, we can also call TB-space **Concept Fock Space** (CFS). The interesting example is the point representing the *vocabulary*, or all distinct terms, of the collection ([2], page 174):

$$|V\rangle = |1,1,1,\ldots,1\rangle \tag{1.3.2}$$

The base vector (1.2.3) now can also be written as:

$$|k_i\rangle = |0,0,\ldots,1,\ldots,0\rangle \equiv |\vec{l}_i\rangle \equiv |(0,0,\ldots,1,\ldots,0)\rangle \tag{1.3.3}$$

They form a subset of the t terms, each represent one term. Here we introduce the t-plet notation of the t base vectors (points) in the Fock space:

$$\vec{l}_i = (0,0,\ldots,1,\ldots,0) \tag{1.3.4a}$$

Note that, in this notation, the addition of two t-plets is actually their union. For example, in our 5-term space, a document vector $|d_1\rangle$ can be written as:

$$|d_1\rangle = |1,1,0,0,1\rangle = |\vec{l}_1 \cup \vec{l}_2 \cup \vec{l}_5 \cup\rangle \tag{1.3.4b}$$

If we allow the term number to be any non-negative integer (or natural number), like the term-frequency in a document, then we have a new representation of the classic vector space: a **Term Frequency Fock Space,** or TF-space for short, which is similar to ONR for Bosons in Quantum mechanics (e.g., photons, mesons…).

The number vector (1.3.4a) can now be generalized to:

$$\vec{N}_i = (0,0,\ldots,N_i,\ldots,0), \quad N_i = \text{any natural number} \tag{1.3.5}$$

The occupation operator now has such a property:

$$\hat{n}_i |\vec{N}\rangle = \hat{n}_i |N_1, N_2, \ldots, N_t\rangle = N_i |N_1, N_2, \ldots, N_t\rangle \tag{1.3.6}$$

In this representation, the three documents and the query in our *GF-Example* now can be written as:

$$\begin{aligned}
|d_1\rangle &= \prod_{i=1}^{t} |N_{1,i}\rangle_i = \prod_{i=1}^{t} |tf_{1,i}\rangle_i = |1,0,1,0,1,1,1,1,1,0,0\rangle \\
|d_2\rangle &= \prod_{i=1}^{t} |N_{2,i}\rangle_i = \prod_{i=1}^{t} |tf_{2,i}\rangle_i = |1,1,0,1,0,0,1,1,0,2,1\rangle \\
|d_3\rangle &= \prod_{i=1}^{t} |N_{3,i}\rangle_i = \prod_{i=1}^{t} |tf_{3,i}\rangle = |1,1,0,0,0,1,1,1,1,0,1\rangle \\
|q\rangle &= \prod_{i=1}^{t} |N_{q,i}\rangle_i = \prod_{i=1}^{t} |tf_{q,i}\rangle = |0,0,0,0,0,1,0,0,0,1,1\rangle
\end{aligned} \tag{1.3.7a}$$

Using term vectors in Eq. (1.3.3) as our base, we can represent the above document vectors and the query vector in columns as follows:

$$|d_1\rangle = \begin{bmatrix} 1 \\ 0 \\ 1 \\ 0 \\ 1 \\ 1 \\ 1 \\ 1 \\ 1 \\ 0 \\ 0 \end{bmatrix} \quad |d_2\rangle = \begin{bmatrix} 1 \\ 1 \\ 0 \\ 1 \\ 0 \\ 0 \\ 1 \\ 1 \\ 0 \\ 2 \\ 1 \end{bmatrix} \quad |d_3\rangle = \begin{bmatrix} 1 \\ 1 \\ 0 \\ 0 \\ 0 \\ 1 \\ 1 \\ 1 \\ 1 \\ 0 \\ 1 \end{bmatrix} \quad |q\rangle = \begin{bmatrix} 0 \\ 0 \\ 0 \\ 0 \\ 0 \\ 1 \\ 0 \\ 0 \\ 0 \\ 1 \\ 1 \end{bmatrix} \qquad (1.3.7b)$$

Note that, in this notation, the addition of two t-plets is actually the addition of number vectors. For example, in a 5-dimetional term space, a document vector $|d_2\rangle$ can be written as:

$$|d_1\rangle = |1,2,0,0,1\rangle = |\vec{l}_1 + \vec{l}_2 + \vec{l}_2 + \vec{l}_5\rangle \qquad (1.3.8)$$

Now we see that the TF-space is a retrieval-closed space of *bounded t-dimensional discrete vectors*, restricted on the vertices of the unit cubes in our WT-space:

$$|w\rangle = \prod_{i=1}^{t}|w_i\rangle_i, \quad w_i \in \aleph \qquad (1.3.9)$$

In TF-space, we do not use vector addition of documents, but we do need to define inner product, which can be derived from their correspondence in WT space:

$$|w\rangle = \prod_{i=1}^{t}|w_i\rangle_i, \quad |u\rangle = \prod_{i=1}^{t}|u_i\rangle_i, \quad \langle w|u\rangle = \sum_{i=1}^{t} w_i u_i \qquad (1.3.10)$$

We will see that the TF-space and their representations will greatly simplify the description of the so-called *Latent Semantic Indexing Model* (see §3). More about Operators in Fock space is given in Appendix B.

### 1.4. The Classical Probabilistic Model and Term-Query Probability Fock Space

Assume that the term-document and term-query weights are all binary, as in eq. (1.2.1), and assume that $R$ is the set of relevant documents for a query $q$ and $\bar{R}$ is the irrelevant document for the query. Then we may modify the identity matrix in eq. (1.1.1) to including a probability factor with respect to $q$, $R$ and $\overline{R}$ :

$$I_q = \sum_{i=1}^{t} |k_i\rangle P_q(k_i, R, \overline{R}) \langle k_i| \equiv \sum_{i=1}^{t} |k_i\rangle (P_q(k_i, R) - P_q(k_i, \overline{R})) \langle k_i|. \qquad (1.4.1)$$

Here $p_q(k_i, R)$ and $p_q(k_i, \overline{R})$ may be computed using formulas like ([2], page 32):

$$p_q(k_i, R) = \log \frac{P(k_i | R)}{1 - P(k_i | R)}$$
$$p_q(k_i, \overline{R}) = \log \frac{P(k_i | \overline{R})}{1 - P(k_i | \overline{R})} \qquad (1.4.2)$$

where $P(k_i | R)$ is the probability when term $k_i$ is present in a document randomly selected from the set $R$ and $P(k_i | \overline{R})$ is the probability when $k_i$ is resent in a document randomly selected from the set $\overline{R}$. Usually, in the beginning, the two probabilities are assigned as constants for all terms, for example ([2], page 33):

$$P(k_i | R) = 0.5, \quad P(k_i | \overline{R}) = \frac{n_i}{N} \qquad (1.4.3)$$

Then, based on the retrieved documents, one can improve the initial ranking.

The Similarity Coefficient now can be calculated as:

$$SC(d_\mu, q) \sim \langle d_\mu | q \rangle \sim \langle d_\mu | I_q | q \rangle$$
$$= \sum_{i=1}^{t} \langle d_\mu | k_i \rangle P(k_i, R, \overline{R}) \langle k_i | q \rangle = \sum_{i=1}^{t} w_{i,\mu} P(k_i, R, \overline{R}) w_{q,i} \qquad (1.4.4)$$

The reason that here we can use $I_q$, instead of $I$, is that we are limited in the subset of terms in query $q$, and all other terms $k_i$, not contained in $q$, are orthogonal to $|q\rangle$, i.e., $w_{i,q} = 0$. Because this space is restricted to subspace spanned by $q$, we may call this space as *Term-Query Probability Fock Space*, or TQP-space for short.

In many cases, we want to normalize $d_\mu$ and $q$, as in Eq. (1.1.8). Now suppose that the query is related to $m$ terms. Then the TQP –space is an $m$-dimensional subspace of the WT-space as described by Eq, (1.1.9):

$$|v\rangle = \sum_{i=1}^{t} w_i | k_i \rangle, \quad |w_i| \in \{0,1\} \subset \aleph, \, w_i = 0 \text{ if } \langle q | k_i \rangle = 0 \qquad (1.4.5)$$

## 2. The Alternative IR Models

**2.1. The Fuzzy Set Retrieval and Fuzzy Boolean Term Fock Space**

In classical Boolean Model, a document either has a term or does not have the term. This fact is represented in the values 0 or 1 in the vectors of (1.2.6a). A document is either relevant or irrelevant to a query. This fact is reflected in the set expression (1.2.6c). The basic assumption can be seen in Eq. (1.2.2b), that is, all term states are eigenstates of occupation operator.

An interesting case in QM is that if a spin state is a linear combination of the up and down status. Then this state is not an eigenvector of $\hat{s}_z$ anymore. We can only predict its "average" value (assume the state vector is normalized):

$$\hat{S}_z = \langle \hat{S}_z \rangle = \langle \psi_s | \hat{S}_z | \psi_s \rangle \tag{2.1.1}$$

Because $\hat{s}_z$ can have eigenvalue only -½ or ½, its average value in the range of [-½, ½]. Based on Eq. (1.2.3c), we can think that if there is dependence between terms, we may assign a fuzzy number to occupation operator for i-th term in the $\beta$-th document as (assume the document vector is normalized):

$$\mu_{i,\beta} \equiv \langle d_\beta | \hat{n}_i | d_\beta \rangle \tag{2.1.2}$$

Because $\hat{n}_i$ can have eigenvalues only 0 or 1, its average value in the range of [0, 1].

This is very mush similar to *fuzzy set theory*, (see [2], page 35 and its Ref. [846]), where we can have *a degree of membership* for each term in a document. This can be done by using a map in our 2D space (remember, each 2D space is for one term):

$$\hat{F}_j | w_{i,v} \rangle_i = | \mu_{i,v} \rangle_i, \quad \mu_{i,v} \in [0,1] \tag{2.1.3}$$

For a document vector, we have the following mapped *fuzzy document vector*:

$$| D_\beta \rangle = \prod_{i=1}^{t} \hat{F}_\beta | w_{i,\beta} \rangle_i = \prod_{i=1}^{t} | \mu_{i,\beta} \rangle_i = | \mu_{1,\beta}, \mu_{2,\beta}, \ldots \mu_{3,\beta} \rangle \tag{2.1.4}$$

The *membership of i-th term in j-th document* $\mu_{i,j}$, can be calculated, e.g., using following formula (see [3], page 86 and our notation in Eq. (1.1.3)):

$$\mu_{i,\beta} = \langle d_\beta | \hat{n}_i | d_\beta \rangle = \frac{tf_{i,\beta}}{dl_\beta} \tag{2.1.5}$$

In our *GF-Example* in §1.1.1, we have:

$dl_1 = 7, \quad dl2_1 = 8, \quad dl_3 = 7$.

Hence, the term gold has a membership in first document as:

$$\mu_{6,1} = \frac{tf_{6,1}}{dl_1} = \frac{1}{7} = 0.143$$

Other memberships are listed in table 2.1 9(see [3], page 86).

Table 2.1: Term Membership in Documents ($\mu_{i,\beta}$)

| Term $i$ | 1 | 2 | 3 | 4 | 5 | 6 | 7 | 8 | 9 | 10 | 11 |
|---|---|---|---|---|---|---|---|---|---|---|---|
| Word | a | arrived | damaged | delivery | fire | *gold* | in | of | shipment | *silver* | *truck* |
| $\mu_{i,1}$ | 0.143 | 0 | 0.143 | 0 | 0.143 | 0.143 | 0.143 | 0.143 | 0.143 | 0 | 0 |
| $\mu_{i,2}$ | 0.125 | 0.125 | 0 | 0.125 | 0 | 0 | 0.125 | 0.125 | 0 | 0.25 | 0.125 |
| $\mu_{i,3}$ | 0.143 | 0.143 | 0 | 0 | 0 | 0.143 | 0.143 | 0.143 | 0.143 | 0 | 0.143 |

Using our notation, the fuzzy document vectors can be written as:

$$|D_1\rangle = \prod_{i=1}^{t} |\mu_{1,j}\rangle_i = |0.143, 0, 0.143, 0, 0.143, 0.143, 0.143, 0.143, 0.143, 0, 0\rangle$$

$$|D_2\rangle = \prod_{i=1}^{t} |\mu_{2,j}\rangle_i = |0.125, 0.125, 0, 0.125, 0, 0, 0.125, 0.125, 0, 0.25, 0.125\rangle$$

$$|D_3\rangle = \prod_{i=1}^{t} |\mu_{3,j}\rangle_i = |0.143, 0.143, 0, 0, 0, 0.143, 0.143, 0.143, 0.143, 0, 0.143\rangle$$

We call this as a *Fuzzy Boolean Term Fock Space*, or FBT-space, which can be thought as an extension of TF-ONR by mapping Frequency to a real number between 0 and 1. Again, it is a retrieval-closed space of vectors.

To get the membership of *i-th* term in *j-th* document, we can define a new *fuzzy occupation operator* $\hat{\mu}_i$ and a *Fuzzy Membership function* $M_i(d_\beta)$:

$$\hat{\mu}_i |D_\beta\rangle = \hat{\mu}_i |\mu_{1,\beta}, \mu_{2,\beta}, \ldots, \mu_{t,\beta}\rangle \equiv \mu_{i,\beta} |D_\beta\rangle$$

$$M_i(d_\beta) \equiv \frac{\langle D_\beta | \hat{\mu}_i | D_\beta \rangle}{\langle D_\beta | D_\beta \rangle} = \mu_{i,\beta} \tag{2.1.4}$$

Note that the *M*-function should satisfy the fuzzy set operation rules ([2], page 35 and [3], page 85):

$$M_{i \cup k}(d_\beta) = \max(\mu_{i,\beta}, \mu_{k,\beta})$$
$$M_{i \cap k}(d_\beta) = \min(\mu_{i,\beta}, \mu_{k,\beta}) \qquad (2.1.5)$$
$$M_{\bar{i}}(d_\beta) = 1 - \mu_{i,\beta}$$

Now let us calculate the membership of documents relative to the following query ([3], page 87):

$$q = (gold \vee silver) \wedge truck = (k_6 \vee k_{10}) \wedge k_{11}$$
$$M_q(d_1) = M_{(6 \cup 10) \cap 11}(d_1) = \min(\max(\mu_{6,1}, \mu_{10,1}), \mu_{11,1}) = \min(\max(0.143, 0.143), 0) = 0$$
$$M_q(d_2) = M_{(6 \cup 10) \cap 11}(d_2) = \min(\max(\mu_{6,2}, \mu_{10,2}), \mu_{11,2}) = \min(\max(0, 0.25), 0.125) = 0.125$$
$$M_q(d_3) = M_{(6 \cup 10) \cap 11}(d_3) = \min(\max(\mu_{6,3}, \mu_{10,3}), \mu_{11,3}) = \min(\max(0.143, 0), 0.143) = 0.143$$

A more explicit way to consider term-term correlation is to introduce a correlation matrix (see [2], page 36 and its Ref. [616]). It can be represented by a matrix representation of a correlation operator:

$$c_{i,l} \equiv \langle k_i | \hat{c} | k_l \rangle = \frac{n_{i,l}}{n_i + n_l - n_{i,l}} \qquad (2.1.6a)$$

Where $n_i$ is defined in Eq. (1.1.3b), and $n_{i,l}$ is the documents which contains both terms. Then the membership of *i-th* term in $\beta$-th document can be computed as:

$$\mu_{i\beta} = \langle d_\beta | \hat{n}_i | d_\beta \rangle = M_i(d_\beta) \equiv 1 - \prod_l \{(1 - c_{i,l}) | tf_{l,\beta} > 0\} \qquad (2.1.6b)$$

In this approach, the fuzzy OR operation defined in (2.1.5) is replaced by an algebraic sum, implemented as a complements of a negative algebraic products, and the fuzzy AND operation in (2.1.5) is computed using algebraic product. For example:

$$M_{a \cup b \cup c}(d_\beta) = 1 - (1 - \mu_{a,\beta})(1 - \mu_{b,\beta})(1 - \mu_{c,\beta})$$
$$M_{a \cap b \cap c}(d_\beta) = \mu_{a,\beta} \mu_{b,\beta} \mu_{c,\beta} \qquad (2.1.7)$$
$$M_{\bar{a}}(d_\beta) = 1 - \mu_{a,\beta}$$

To use above set operation rules, one should first decompose the query concepts in a *distinctive normal form*, as a union of disjunctive concept vectors (see [2], pages 36-37).

In any case, it seems that, using FBT-space can greatly simplify the expressions and calculations for fuzzy set examples. Based on our discussion, FBT-space is a *retrieval-closed space of bounded t-dimensional continuous vectors*, as described by Eq. (1.1.9).

**2.2. Extended Boolean Model**

Another way to improve Boolean search is to rank the relevance document based on its closeness to a query, using Extended Boolean Model, introduced by Salton, Fox and Wu [7, 8] (see [2], page 38, [3], page 67 and [5]).

First, we assume that all terms in a document or a query are assigned with weights as in classical vector model, described in §1.1. All weights are normalized to be in the range of [0, 1].

Next, the least or most desirable point is decided based on the operation of the Boolean query (union or join). If the query is a union of *m* terms, we will have the least desirable point; if the query is a joint of m terms, then we will have the most desirable points. To describe such points, we restrict ourselves to a subspace, spanned by the terms contained in the query. We call this space the *q*-space (like in the classical probabilistic model). To simplify our formula, we assume that *the first m terms are contained in the query*. Now we can express the Identity operator in the *m*-dimensional *q*-space by:

$$\hat{I}_q = \sum_{i=1}^{m} |k_i\rangle\langle k_i| \qquad (2.2.1)$$

In this subspace, the query and a document is expressed in the WT-space (Weighted Term Vector Space**)** as:

$$|q\rangle_q = \hat{I}_q |q\rangle = \sum_{i=1}^{m} |k_i\rangle\langle k_i|q\rangle = \sum_{i=1}^{m} w_{i,q} |k_i\rangle; \quad |d_\beta\rangle_q = \sum_{i=1}^{m} w_{i,\beta} |k_i\rangle. \qquad (2.2.2)$$

If the query is a union of *m* terms,

$$q_\vee = k_1 \vee k_2 \vee \ldots \vee k_m \qquad (2.2.3a)$$

then the *least desirable point* is the origin of the weighted space ([2], page 39):

$$|\vec{0}\rangle_q = \sum_{i=1}^{m} 0 \cdot |k_i\rangle = 0; \qquad (2.2.3b)$$

The same vector, if represented in our TF-space (or BT-space), will be:

$$|\vec{0}\rangle_q = |0,0,0,\ldots,0\rangle \qquad (2.2.3c)$$

If the query is a joint of *m* terms:

$$q_\wedge = k_1 \wedge k_2 \wedge \ldots \wedge k_m \tag{2.2.3d}$$

We can see that the *most desirable point* is the vector of the highest weights ([2], page 39):

$$|\vec{1}\rangle_q = \sum_{i=1}^{m} 1 \cdot |k_i\rangle = \sum_{i=1}^{m} |k_i\rangle; \tag{2.2.3e}$$

The same vector, if represented in our TF-space (or BT-space), will be:

$$|\vec{1}\rangle_q = |1,1,1,\ldots,1\rangle \tag{2.2.3f}$$

Now we need define a measure or a distance in the space, in order to calculate the distance of the document from such point. According to [7.8], we can define the normalized distance between two vectors in the m-dimensional *q*-space with a parameter *p* by:

$$d^p(|a\rangle_q, |b\rangle_q) = \left[ \frac{\sum_{i=1}^{m} |w_{i,a} - w_{i,b}|^p}{m} \right]^{1/p} \tag{2.2.4}$$

This is a normalized form of the *Minkowski distance of order p* (or *p-norm distance*) in a Euclidean space, as an example of metric spaces [9].

With the above descriptions, we can simplify the Similarity Coefficients (*SC*) formulas as follows.

$$SC(q_\vee, d_\beta) = d^p(|d_\beta\rangle_q, |\vec{0}\rangle_q) \tag{2.2.5a}$$

$$SC(q_\wedge, d_\beta) = 1 - d^p(|\vec{1}\rangle_q - |d_\beta\rangle_q) \tag{2.2.5b}$$

If $p = 1$, the distance is the difference of the two vectors in *m*-dimensional vector space; if $p = 2$, the distance is for an *m*-dimensional Euclidian space; if $p = \infty$, Eq. (2.25) return the result of the original Boolean query. We can also include the weights of query in the calculation of distance ([3], page 68).

We call the space the Extended Boolean Term Fock Space (EBT-space). Based on our discussion, the EBT-space is *a bounded t-dimensional continuous metric space*: it is a retrieval-closed space of vectors as described by Eq. (1.1.9) and it is also a metric space with a *p*-norm (2.2.4) or (2.2.6) as its distance function. No inner product is need here.

## 2.3. Term-Term Correlation and the Document Fock Space

Because the existence of term-term correlations, we cannot say terms are independent. This means that the term vectors in Eq. (1.1.b) are not orthogonal to each other anymore:

$$\langle k_i | k_j \rangle = c_{ij} \neq \delta_{ij} .\quad (2.3.1a)$$

Note that the completeness of terms, as described in Eq. (1.1.1c) still holds as long as the expansion of documents and query are unique with respect to defined inner products. In this case, we still have:

$$\sum_{i=1}^{t} | k_i \rangle \langle k_i | = I. \quad (2.3.1b)$$

There are many ways to compute term-term relationships. The *Generalized Vector Model* was proposed by Wong, Ziarko and Wong in 1985 [6]. They considered the space spanned by the $2^t$ vectors similar those described in Eq. (1.2.6), introduced $2^t$ orthonormal vectors as the bases of the vector space, expressed terms as linear combinations of these base vectors, and calculated term-term correlation based on such term expressions. More details can also be seen in [2: pages 41-44].

Here we discuss a query expansion model based on a global similarity thesaurus which is constructed automatically ([10], and [2], page 131-133). Because the conditions in Eq. (2.3.1), we cannot use the base vectors in (1.1.1b) to represent our vector space. Instead, we expand term vectors with respect to documents:

$$| k_i \rangle = \sum_{\beta=1}^{N} | d_\beta \rangle \langle d_\beta | k_i \rangle \equiv \sum_{i=1}^{N} w_{i,\beta} | d_\beta \rangle \quad (2.3.2)$$

Assume any term is at least in one document and no two documents are identical in term frequency, so term vector can be uniquely expanded with respect to document vectors. This is a very strong requirement: the Term-Document matrix defined in Eq. (3.1.1) need have a rank of *t*. In this way, we can think that the document vectors form a basis of N-dimensional Fock space, with the inner product of term-document defined in (2.3.2):

$$| d_\alpha \rangle : \quad \langle d_\alpha | = (| d_\alpha \rangle)^T ; \quad \sum_{\beta=1}^{N} | d_\beta \rangle \langle d_\beta | = I_d \quad (2.3.3)$$

The weights in Eq. (2.3.2), now the document occupation number of terms, are calculated using the following formula ([2], page 132),

$$w_{i,\beta} = \frac{\left[0.5 + 0.5\left(tf_{i,\beta} / \max_{\beta}(tf_{i,\beta})\right)\right] \cdot itf_{i,\beta}}{\sqrt{\sum_{\mu=1}^{N}\left[0.5 + 0.5\left(tf_{i,\mu} / \max_{\mu}(tf_{i,\mu})\right)\right]^2 \cdot itf_{i,\mu}^2}} \quad (2.3.4)$$

where $tf_{i,\mu}$ is defined in Eq. (1.1.3a), $\max_{\mu}(tf_{i,\mu})$ is the maximum value of all $tf_{i,\mu}$ for a given $|k_i\rangle$, and $itf_{i,\mu}$ is the inverse term frequency for document $|d_\beta\rangle$, defined by:

$$itf_\beta = \log \frac{t}{t_\beta} \quad (2.3.5)$$

Here $t_\beta$ is the number of distinct terms in document $d_\beta$.

Now we can express the term-term correlation factor as (identical to Eq. (5.12) in [2], page 132):

$$c_{\mu,\nu} = \langle k_\mu | k_\nu \rangle = \langle k_\mu | I_d | k_\nu \rangle = \sum_{\beta=1}^{N} \langle k_\mu | d_\beta \rangle \langle d_\beta | k_\nu \rangle = \sum_{\beta=1}^{N} w_{\mu,\beta} w_{\nu,\beta} \quad (2.3.6)$$

Although terms are not an orthonormal system, we still can define the query vector as a linear combination of terms contained in the query ([2], page 133):

$$|q\rangle = \sum_{k_i \in q} w_{i,q} |k_i\rangle, \quad \langle q| = \sum_{k_i \in q} w_{i,q} \langle k_i| \quad (2.3.7)$$

Here, the weights are calculated similarly to Eq. (2.3.5), but they are not equal to the inner product of term-query. Instead, the inner product of a query with a term is computed using (2.3.7) and (2.3.6) as:

$$\langle q | k_\nu \rangle = \sum_{k_{ui} \in q} w_{u,q} \langle k_u | k_\nu \rangle = \sum_{k_{ui} \in q} w_{u,q} c_{u,\nu} \quad (2.3.8)$$

Now we can derive our expression for the Similarity Coefficient as:

$$SC(q, d_\beta) = \langle q | d_\beta \rangle = \sum_{k_u \in q} \langle q | k_u \rangle \langle k_\nu | d_\beta \rangle = \sum_{k_\nu \in q} w_{\nu,q} c_{u,\nu} w_{u,\beta} \quad (2.3.9)$$

This equation is identical to the one on page 134, Ref [2].

We call the vector space spanned by document vectors defined in (2.3.3) as Document Vector Space (DV-space), which is an N-dimensional vector space with base vector (2.3.3), and its inners product of document and query with term as (2.3.2) and (2.3.8). Based on above rules, we can then calculate the term-term correlations as in (2.3.6).

We see in this model, document and query are treated differently, because they have different inner product rules with terms. In next section, we will see a more natural way of introduce term-term interaction in which document and query would be treated in the same way, unified by a metric tensor derived using SVD.

## 3. Latent Semantic Indexing Model and SVD

In WT-space of classical vector models, we have the basic assumption that the term vectors form a complete orthonormal bases as in Eq. (1.1.1b). In reality, terms are not independent vectors, because of the existence of the term-document relations.

This situation is very much like the situation in the quantum system of the Hydrogen atom when neglecting the orbital-spin (*L-S*) interaction ([1], pages 549-554). If we neglect the *L-S* interaction, the Hamiltonian (*H*) of the system has eigenvector which are also the common eigenvectors of the orbital angular momentum operators ($L^2$, $L_z$) and the spin angular momentum ($S^2$, $S_z$). Any two of them are orthogonal if anyone of the four eigenvalues ($L^2$, $L_z$, $S^2$, $S_z$) is different. But because of the *L-S* interaction, they become dependent and we have no such common eigenvectors anymore. Instead, we have to introduce the total angular momentum (***J*** = ***L*** + ***S***), to find the common eigenvectors of ($J^2$, $L^2$, $S^2$, $J_z$), which give us the fine structure of the Hydrogen atom ([1], page 554-555).

### 3.1. Term-Document, Term-Term and Document-Document Arrays

In *Latent Semantic Indexing* (LSI) Model *singular value decomposition* (SVD), the key assumption is that the document-term dependence causes both term-term and documents-document dependences ([2], pages 44-45; [3], pages 70-73; [11-12]). The document-term dependence is nothing else but the frequency of *i-th* term in the $\mu$ *-th* document, $tf_{i,\mu}$, as defined in Eq. (1.1.3a). We also have their representations in Eq. (1.3.7) in our *TF*-space for the *GF*-examples. This relation can be viewed as the definition of inner product between a document and a term, and represented by a *t x N* matrix *A* (the *term-document matrix*) as follows:

$$A_{i,\beta} = \langle k_i | d_\beta \rangle = \langle d_\beta | k_i \rangle = tf_{i,\beta}$$
$$A = [| d_1 \rangle | d_2 \rangle ... | d_N \rangle] \tag{3.1.1}$$

Here we have used the document vectors as the columns in the matrix. From now on, we assume that *t > N*. Apply to our GF-example, we have the following three document vectors, the term-doc matrix and the query vector.

```
Terms           d1  d2  d2       q
  ↓             ↓   ↓   ↓        ↓

a               ⎡1   1   1⎤     ⎡0⎤
arrived         ⎢0   1   1⎥     ⎢0⎥
damaged         ⎢1   0   0⎥     ⎢0⎥
delivery        ⎢0   1   0⎥     ⎢0⎥
fire       A = ⎢1   0   0⎥  q = ⎢0⎥
gold            ⎢1   0   1⎥     ⎢1⎥
in              ⎢1   1   1⎥     ⎢0⎥
of              ⎢1   1   1⎥     ⎢0⎥
shipment        ⎢1   0   1⎥     ⎢0⎥
silver          ⎢0   2   0⎥     ⎢1⎥
truck           ⎣0   1   1⎦     ⎣1⎦
```

**[12]-4: Figure 2. Term-document matrix and query matrix example**

The term-term interaction is defined by a *t x t* matrix *L* (called *Left matrix*):

$$L_{i,j} = (AA^T)_{i,j} = \sum_\beta \langle k_i | d_\beta \rangle \langle d_\beta | k_j \rangle = \sum_\beta tf_{i,\beta}\, tf_{j,\beta}$$

$$L = AA^T = [|d_1\rangle |d_2\rangle ... |d_N\rangle] \begin{bmatrix} \langle d_1 | \\ \vdots \\ \langle d_N | \end{bmatrix}$$

(3.1.2)

Here we have used the transpose of document vectors as the rows in the matrix. Note that the elements of matrix *L* are similar to (2.3.6); the difference is the definition of the inner product of term and document.

The document-document interaction is defined by *N x N* matrix *R* (called *Right matrix*):

$$R_{\alpha,\beta} = (A^T A)_{\alpha,\beta} = \sum_i \langle d_\alpha | k_i \rangle \langle k_i | d_\beta \rangle = \sum_i tf_{i,\alpha}\, tf_{i,\beta}$$

$$R = A^T A = \begin{bmatrix} \langle d_1 | \\ \vdots \\ \langle d_N | \end{bmatrix} [|d_1\rangle |d_2\rangle ... |d_N\rangle]$$

(3.1.3)

Because both *L* and *R* are real symmetric matrices, they both have real, complete and orthogonal eigenvectors with the same set of real, non-negative eigenvalues ([1], pages 199-204; [13], pages 321-325). Here are their normalized forms:

$$L|u_i\rangle = \lambda_i |u_i\rangle, \quad \langle u_i | u_j \rangle = \delta_{ij}$$
$$R|v_\beta\rangle = \lambda_\beta |v_\beta\rangle, \quad \langle v_\alpha | v_\beta \rangle = \delta_{\alpha\beta}$$

(3.1.4)

## 3.2. SVD, Vector Inner Product and Metric Tensor

Now we can apply the Singular Value Decomposition (SVD) of matrix A to use the eigenvectors of both *L* and *R*:

$$A = U \, S \, V^T \tag{3.2.1}$$

Here, *U* and *V* are orthogonal matrices, the columns of the *t x N* matrix *U* are the *N* new term eigenvectors **u**, the columns of *N x N* matrix *V* are the new document eigenvectors **v**, and the *N x N* matrix S is a diagonal matrix with its diagonal elements $S_i = \sqrt{\lambda}$ (*the singular values*) as the square root of the eigenvalues of *M* or $M^T$. We can arrange $S_i$ such that $S_1 > S_2 > S_3 > \ldots S_N$.

One of the advantages of SVD is that, in most cases, we do not need to use eigenvectors of zero eigenvalues (usually, there are many of them), and we can also keep only *r* non-zero singular values to reduce the ranks of all matrices to *r,* without losing much of the accuracy:

$$A_r = U_r S_r V_r^T \tag{3.2.2}$$

This is referred to *Reduced* SVD (RSVD). We can visually explain the reduction by the following figure (a copy of Figure 8, tutorial 3, Ref. [12], where *k* is used as our *r*).

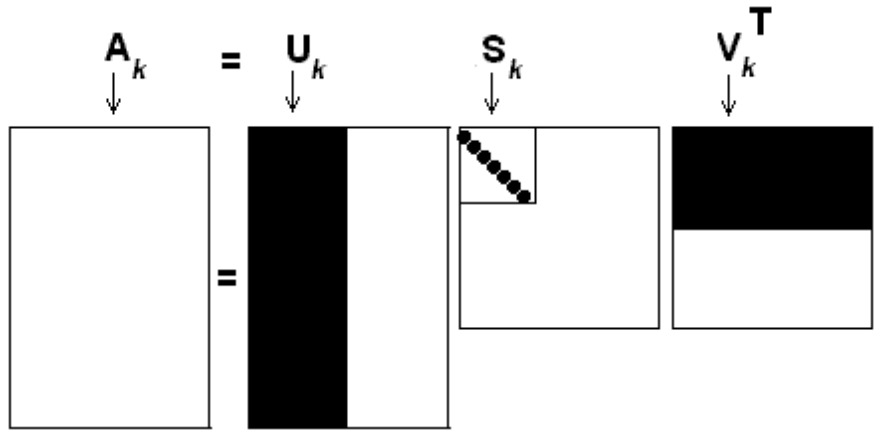

**[12]-4: Figure 8. The Reduced SVD or Rank *k* Approximation**

Using the Dirac notation, we can express (3.2.2) as:

$$[|d_1\rangle |d_2\rangle \ldots |d_N\rangle] = [|k'_1\rangle |k'_2\rangle \ldots |k'_r\rangle] \begin{bmatrix} S_1 & & \\ & \ddots & \\ & & S_r \end{bmatrix} [|d'_1\rangle |d'_2\rangle \ldots |d'_N\rangle] \tag{3.2.3}$$

Here we have used following notations:

$$U = [|u_1\rangle | u_2\rangle ... | u_r\rangle] = [|k'_1\rangle | k'_2\rangle ... | k'_r\rangle]$$
$$V^T = [|v_1\rangle_r | v_2\rangle_r ... | v_r\rangle_r]^T = [|d'_1\rangle | d'_2\rangle ... | d'_N\rangle] \quad (3.2.4a)$$

This means that:

$$V = \begin{bmatrix} \langle d'_1| \\ \vdots \\ \langle d'_N| \end{bmatrix} \quad (3.2.4b)$$

Also note that here the new document vectors are reduced to *r*-dimensional.

A very important equation derived from Eq. (3.2.1) is ([12], tutorial 4):

$$V = A^T U S^{-1} \quad (3.2.5a)$$

Using Eq. (3.1.1) and (3.2.4a), we can express (3.2.5a) in reduced form as:

$$\begin{bmatrix} \langle d'_1| \\ \vdots \\ \langle d'_N| \end{bmatrix} = \begin{bmatrix} \langle d_1| \\ \vdots \\ \langle d_N| \end{bmatrix} [|k'_1\rangle | k'_2\rangle ... | k'_r\rangle] \begin{bmatrix} 1/s_1 & & \\ & \ddots & \\ & & 1/s_r \end{bmatrix} \quad (3.2.5b)$$

This equation tells us how we can transform original document vector $\langle d_\beta|$ to new vector: $\langle d'_\beta|$:

$$\langle d'_\beta| = \langle d_\beta| [|k'_1\rangle | k'_2\rangle ... | k'_r\rangle] \begin{bmatrix} 1/s_1 & & \\ & \ddots & \\ & & 1/s_1 \end{bmatrix}$$

$$\langle d'_\beta| = \begin{bmatrix} \frac{1}{s_1}\langle d_\beta|k'_1\rangle & \cdots & \frac{1}{s_k}\langle d_\beta|k'_r\rangle \end{bmatrix} \quad (3.2.6a)$$

Since the query vector is treated like a document vector, we can compute the new query vector as:

$$\langle q'| = \begin{bmatrix} \frac{1}{s_1}\langle q|k'_1\rangle & \cdots & \frac{1}{s_r}\langle q|k'_r\rangle \end{bmatrix}, \quad |q'\rangle = \begin{bmatrix} \frac{1}{s_1}\langle k'_1|q\rangle \\ \vdots \\ \frac{1}{s_r}\langle k'_r|q\rangle \end{bmatrix} \quad (3.2.6b)$$

Eq. (3.2.6) give us a geometrical interpretation of the transformation: the new document vector is represented in a new base of the new term vectors ($|k'_i\rangle$, eigenvectors of *L*); its $i^{th}$ component is its inner product with corresponding new term vector $|k'_i\rangle$, scaled by a factor of $1/S_i$. So we can say the new document is the result of a rotation (from ***k*** to ***k'***) plus a scaling. Then, the inner product of the two new vectors can be derived from Eq. (3.2.6) as:

$$\langle d'_\beta | q' \rangle = \sum_{a=1}^{r} \frac{1}{s_a^2} \langle d_\beta | k'_a \rangle \langle k'_a | q \rangle \quad (3.2.7)$$

Finally, the ranking of documents now can be calculated using the Cosine low (1.1.5), or:

$$SC(d_\mu, q) = \frac{\langle d'_\mu | q' \rangle}{|d'_\mu| * |q'|} = \frac{\langle d'_\mu | q' \rangle}{\sqrt{\langle d'_\mu | d'_\mu \rangle} * \sqrt{\langle q' | q' \rangle}} \quad (3.2.10)$$

Now let us introduce a metric tensor to simplify our calculations. We can rewrite Eq. (3.2.7) in a more concise form:

$$\langle d'_\beta | q' \rangle = \langle d_\beta | \left( \sum_{a=1}^{r} \frac{1}{s_a^2} | k'_a \rangle \langle k'_a | \right) | q \rangle = d_{\beta,j} g_{jl} q_l \quad (3.2.11)$$

Here, we have defined our Riemannian metric tensor as:

$$\hat{g} = \sum_{a=1}^{r} \frac{1}{s_a^2} |k'_a\rangle\langle k'_a|, \quad g_{jl} = \langle k_j | \hat{g} | k_l \rangle = \sum_{a=1}^{r} \frac{1}{s_a^2} k'_{a,j} k'_{a,l} \quad (3.2.12)$$

In this way, we actually interpret the inner product of two vectors on the tangent space of an *r*-dimensional Riemannian Manifold [16][19]:

$$\langle d_\beta, q \rangle \equiv \langle d'_\beta | q' \rangle = \langle d_\beta | \hat{g} | q \rangle = \sum_{j,l=1}^{t} g_{jl} d_{\beta,j} q_l \quad (3.2.13)$$

The norm of the vector can be calculated as the square root of its self- inner product. Eq. (3.2.10) now can be rewritten as:

$$SC(d_\mu, q) = \frac{\langle d_\mu | \hat{g} | q \rangle}{\sqrt{\langle d_\mu | \hat{g} | d_\mu \rangle} * \sqrt{\langle q | \hat{g} | q \rangle}} \quad (3.2.14)$$

### 3.3. Term-Frequency Fock Space and Metric Tensor for GF-example

We see that, from Eqs. (3.2.11-13), to compute the ranking of a document $d_\mu$ with respect to a query *q*, all we need to do are:

1. Calculate the term-term interaction matrix $L$ from term-document frequency matrix $A$, using Eq. (3.1.2).
2. Calculate the eigenvalues ($\lambda_i = s_i^2$) of $L$, and then keep eigenvectors ($|k'_i\rangle$) for top $r$ non-zero eigenvalues ($S_1 > S_2 > \ldots > S_r > \ldots > 0$).
3. Calculate the $t \times t$ metric tensor $g$ using Eq. (3.2.12).
4. Calculate the ranking using (3.2.14).

Now let us go over it with our GF-example.

**Step 1**: we use online tool [14] to calculate $L = AA^T$, the matrix is as follows:

$$\begin{matrix}
3 & 2 & 1 & 1 & 1 & 2 & 3 & 3 & 2 & 2 & 2 \\
2 & 2 & 0 & 1 & 0 & 1 & 2 & 2 & 1 & 2 & 2 \\
1 & 0 & 1 & 0 & 1 & 1 & 1 & 1 & 1 & 0 & 0 \\
1 & 1 & 0 & 1 & 0 & 0 & 1 & 1 & 0 & 2 & 1 \\
1 & 0 & 1 & 0 & 1 & 1 & 1 & 1 & 1 & 0 & 0 \\
2 & 1 & 1 & 0 & 1 & 2 & 2 & 2 & 2 & 0 & 1 \\
3 & 2 & 1 & 1 & 1 & 2 & 3 & 3 & 2 & 2 & 2 \\
3 & 2 & 1 & 1 & 1 & 2 & 3 & 3 & 2 & 2 & 2 \\
2 & 1 & 1 & 0 & 1 & 2 & 2 & 2 & 2 & 0 & 1 \\
2 & 2 & 0 & 2 & 0 & 0 & 2 & 2 & 0 & 4 & 2 \\
2 & 2 & 0 & 1 & 0 & 1 & 2 & 2 & 1 & 2 & 2
\end{matrix}$$

**Step 2**: we use online tool [15] to calculate its eigenvalues and eigenvectors. It has 3 non-zero eigenvalues, their square roots $S_i$ are:

$$\{S_1, S_2, S_3\} = \{4.0989, 2.3616, 1.2737\} \qquad (3.3.1)$$

Their corresponding eigenvectors ($k'_1, k'_2, k'_3$) are as follow:

[-0.4201, -0.2995, -0.1206, -0.1576, -0.1206, -0.2626, -0.4201, -0.4201, -0.2626, -0.3151, -0.2995]$^T$
[-0.0748, 0.2001, -0.2749, 0.3046, -0.2749, -0.3794, -0.0748, -0.0748, -0.3794, 0.6093, 0.2001]$^T$
[-0.0460, 0.4078, -0.4538, -0.2006, -0.4538, 0.1547, -0.0460, -0.0460, 0.1547, -0.4013, 0.4078]$^T$

**Step 3**: we keep all three eigenvectors, i.e., set $r = 3$ in Eq. (3.2.2); using a simple computer program [18] and Eq. (3.2.12), we get the metric matrix g as follows:

```
 0.0127  -0.0067  0.0195   0.0055   0.0195   0.0072   0.0127   0.0127   0.0072   0.011   -0.0067
-0.0067   0.1149 -0.1217  -0.0366  -0.1217   0.0299  -0.0067  -0.0067   0.0299  -0.0733  0.1149
 0.0195  -0.1217  0.1412   0.0421   0.1412  -0.0226   0.0195   0.0195  -0.0226   0.0844 -0.1217
 0.0055  -0.0366  0.0421   0.0428   0.0421  -0.0373   0.0055   0.0055  -0.0373   0.0857 -0.0366
 0.0195  -0.1217  0.1412   0.0421   0.1412  -0.0226   0.0195   0.0195  -0.0226   0.0844 -0.1217
 0.0072   0.0299 -0.0226  -0.0373  -0.0226   0.0446   0.0072   0.0072   0.0446  -0.0747  0.0299
 0.0127  -0.0067  0.0195   0.0055   0.0195   0.0072   0.0127   0.0127   0.0072   0.011  -0.0067
 0.0127  -0.0067  0.0195   0.0055   0.0195   0.0072   0.0127   0.0127   0.0072   0.011  -0.0067
 0.0072   0.0299 -0.0226  -0.0373  -0.0226   0.0446   0.0072   0.0072   0.0446  -0.0747  0.0299
 0.011   -0.0733  0.0844   0.0857   0.0844  -0.0747   0.011    0.011   -0.0747   0.1716 -0.0733
-0.0067   0.1149 -0.1217  -0.0366  -0.1217   0.0299  -0.0067  -0.0067   0.0299  -0.0733  0.1149
```

**Step 4**: Using the simple computer programs [18] and Eq. (3.2.10 and12), we get the following similarity coefficients:

$$SC(d_1, q) = <d1, q>/|d1|/|q| = -0.2787$$
$$SC(d_1, q) = <d1, q>/|d1|/|q| = 0.7690 \quad (3.3.2)$$
$$SC(d_1, q) = <d1, q>/|d1|/|q| = 0.5756$$

We see the ranking of documents are:

$$SC(d_2, q): SC(d_3, q): SC(d_1, q) = 1: 0.85: -0.362, \quad d_2 > d_3 > d_1 \quad (3.3.3)$$

To compare with the results in [3] and [12], we keep only top 2 eigenvectors of L, i.e., set $r = 2$ in Eq. (3.2.12) and metric matrix $g$ now becomes

```
 0.0114   0.0047   0.0066  -1.0E-4   0.0066   0.0116   0.0114   0.0114   0.0116  -2.0E-4   0.0047
 0.0047   0.0125  -0.0077   0.0137  -0.0077  -0.0089   0.0047   0.0047  -0.0089   0.0274   0.0125
 0.0066  -0.0077   0.0144  -0.0138   0.0144   0.0205   0.0066   0.0066   0.0205  -0.0277  -0.0077
-1.0E-4   0.0137  -0.0138   0.018   -0.0138  -0.0182  -1.0E-4  -1.0E-4  -0.0182   0.0361   0.0137
 0.0066  -0.0077   0.0144  -0.0138   0.0144   0.0205   0.0066   0.0066   0.0205  -0.0277  -0.0077
 0.0116  -0.0089   0.0205  -0.0182   0.0205   0.0298   0.0116   0.0116   0.0298  -0.0365  -0.0089
 0.0114   0.0047   0.0066  -1.0E-4   0.0066   0.0116   0.0114   0.0114   0.0116  -2.0E-4   0.0047
 0.0114   0.0047   0.0066  -1.0E-4   0.0066   0.0116   0.0114   0.0114   0.0116  -2.0E-4   0.0047
 0.0116  -0.0089   0.0205  -0.0182   0.0205   0.0298   0.0116   0.0116   0.0298  -0.0365  -0.0089
-2.0E-4   0.0274  -0.0277   0.0362  -0.0277  -0.0365  -2.0E-4  -2.0E-4  -0.0365   0.0724   0.0274
 0.0047   0.0125  -0.0077   0.0137  -0.0077  -0.0089   0.0047   0.0047  -0.0089   0.0274   0.0125
```

Using this metric tensor in Eq. (3.2.13-14), we recalculate the similarity coefficients:

$$SC(d_1, q) = <d1, q>/|d1|/|q| = -0.0552$$
$$SC(d_2, q) = <d2, q>/|d2|/|q| = 0.9912 \quad (3.3.4)$$
$$SC(d_3, q) = <d3, q>/|d3|/|q| = 0.4480$$

The results are almost identical to the results in [3] and [12] (page 73 of [3]; tutorial 4, Figure 6 of [12]). Their results are:

$$SC(d_1, q) = -0.0541$$
$$SC(d_2, q) = 0.9910 \quad (3.3.5)$$
$$SC(d_1, q) = 0.4478$$

We see that, although the order of the ranking of documents is the same as in Eq. (3.3.3) where we use r = 3, but the relative difference seem better than (3.3.3):

$$SC(d_2, q): SC(d_3, q): SC(d_1, q) = 1: 0.451: -0.0545, \quad d_2 > d_3 > d_1 \quad (3.3.6)$$

From the numerical results, we see there is an important characteristic of LSI/SVD: the relevance coefficient can be negative. This means, the angle between the document and the query can be greater than $\pi/2$. If we use the cosine squared law as in Eq. (1.1.11) of Ref [19], then the ranking may cause problem.

### 3.4. Geometry of SVD Metric Space

In this section we would like to give a brief geometric interpretation on metric tensor, SVD and TF-space of LSI.

We know that in the Cartesian coordinate system of a Euclidian space, the metric has the simplest form:

$$g_{jl} = \delta_{jl} \tag{3.4.1}$$

But in other curvilinear coordinate systems (like polar or spherical), metric tensors have different forms, and they can be transformed to each other. But the space is still flat (the curvature is zero). In general relativity (see [17], chapter 3) the metric tensor is the solution of Einstein field equation, representing a curved space by the distribution of matter, modular to general coordinate transformations.

Metric tensor for IR has mentioned in Ref [19] (page 86). Following the notation there, we can rewrite Eq. (3.2.12) as:

$$g_{jl} = \sum_{a=1}^{r} \frac{1}{s_a^2} k'_{a,j} k'_{a,l} = \sum_{a=1}^{r} e_{j,a} e_{l,a} = \sum_{a=1}^{r} \langle e_j | a \rangle \langle a | e_l \rangle = \langle e_j | e_l \rangle \tag{3.4.2}$$

Here the *r*-dimensional base vectors have the following components:

$$\langle a | e_j \rangle = e_{j,a} = \frac{1}{s_i} k'_{a,j} \tag{3.4.3}$$

Now Eq. (3.2.13) can be rewrite as:

$$\langle d_\beta, q \rangle \equiv \langle d'_\beta | q' \rangle = \sum_{j,l=1}^{r} \langle e_j | d_{\beta,j} q_l | e_l \rangle = \sum_{j,l=1}^{t} g_{jl} d_{\beta,j} q_l \tag{3.4.4}$$

The following are 2D graphic representations of the related vectors.

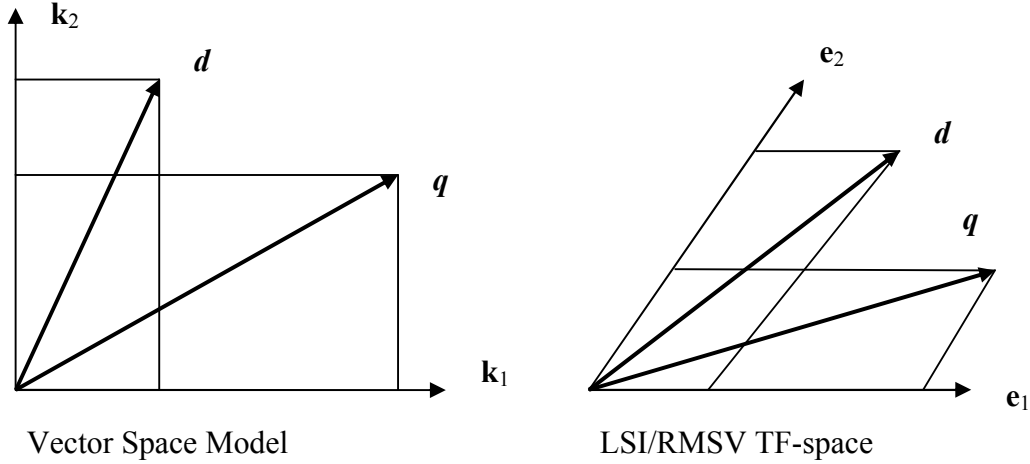

| Vector Space Model | LSI/RMSV TF-space |

In short, we have provided a concrete example of metric tensor from LSI. We see that, because of the term-term interaction, the old term coordinate system (in which we have the representation of term-document interaction array) is no longer a Cartesian coordinated system. *The Term Frequency space of LSI is a flat space*. This is easy to see, because the metric tensor defined in Eq. (3.2.12) or (3.4.2) is a global constant tensor.

### 3.5. SVD Metric Unit-Sphere Map and Document Clustering

We have used Riemannian metric here to define the inner product of vectors on the tangent space. But Riemannian metric can also be used as a distance function. Inspired by the diffusion map in Ref. [20-21], we propose the following SVD unit-sphere map:

1. The document vectors are mapped from the original t-dimensional term-space with base $k_i$ to the t-dimensional space with the new base $e_i$, which are given in Eq. (3.4.3) in terms of the r-eigenvectors of left matrix $L$.

2. Using the metric in Eq. (3.4.2), make each document *a unit vector*, so all documents and queries are mapped onto the *t-1* dimensional unit-sphere:

$$|d_\beta\rangle \mapsto |\upsilon_\beta\rangle \equiv |d_\beta\rangle / \sqrt{\langle d_\beta | \hat{g} | d_\beta \rangle}, \quad |\upsilon_\beta| = 1. \qquad (3.5.1)$$

3. Calculate the *distance between any two data points* on the *unit-sphere*:

$$d(|\upsilon_\alpha\rangle, |\upsilon_\beta\rangle) \equiv \sqrt{(\langle \upsilon_\alpha | - \langle \upsilon_\beta |) \hat{g} (|\upsilon_\alpha\rangle - |\upsilon_\beta\rangle)}$$
$$= \sqrt{2 - 2\langle \upsilon_\alpha | \hat{g} | \upsilon_\beta \rangle} = \sqrt{2 - 2\cos(\theta_{\alpha,\beta})} = 2|\sin(\frac{\theta_{\alpha,\beta}}{2})| \qquad (3.5.2)$$

4. Here $\cos(\theta_{\alpha,\beta})$ can be calculated using the metric defined in Eq. (3.4.2):

$$\cos(\theta_{\alpha,\beta}) = \langle \upsilon_\alpha | \hat{g} | \upsilon_\beta \rangle = \frac{\langle d_\alpha | \hat{g} | d_\beta \rangle}{\sqrt{\langle d_\alpha | \hat{g} | d_\alpha \rangle \langle d_\beta | \hat{g} | d_\beta \rangle}} \qquad (3.5.3)$$

Applying to GF-example, using [18] with $r = 2$ as for Eq. (3.3.5), we have the following distances between docs and the query:

|       | $q$    | $d_1$  | $d_2$  | $d_3$  |
|-------|--------|--------|--------|--------|
| $q$   | 0      | 1.4547 | 0.1326 | 1.0507 |
| $d_1$ | 1.4547 | 0      | 1.5422 | 0.5140 |
| $d_2$ | 0.1326 | 1.5422 | 0      | 1.1638 |
| $d_3$ | 1.0507 | 0.5140 | 1.1638 | 0      |

We see the order of closeness of documents to $q$ is as in Eq. (3.3.6):

$d_2 > d_3 > d_1$

Thus, we have a way to define the *neighborhood* of data points. By adjusting the *radius of the neighborhood* (RON), we can group points with intersecting neighborhood as a cluster. If we set the RON = 0.52 in our example, we see that $q$ and $d_2$ are in one group, while $d_3$ and $d_1$ are in another group. If we set RON = 0.2, then we have only one group ($q$ and $d_2$).

If we use [18] with $r = 3$ as for Eq. (3.3.2), the results are as follows. Again, we see that the relative closeness would be better if we use r = 2.

|       | $q$    | $d_1$  | $d_2$  | $d_3$  |
|-------|--------|--------|--------|--------|
| $q$   | 0      | 1.5991 | 0.6797 | 0.9213 |
| $d_1$ | 1.5991 | 0      | 1.4156 | 1.4154 |
| $d_2$ | 0.6797 | 1.5422 | 0      | 1.4149 |
| $d_3$ | 0.9213 | 1.4151 | 1.4149 | 0      |

**4. Summary**

We have *used Dirac notation* and *introduced Fock Spaces* to unify the representations of IR models. The classical Boolean model can be described by **Concept Fock Space** or Boolean Term Fock Space, with occupation number 0 or 1. If we relax the restrictions on the value of occupation number, we can define **Fuzzy Boolean Term Fock space** (for fuzzy Boolean model), **Term Frequency Fock space** (For LSI), and Weighted Term Fock Space (for classical vector model). The term vectors in such a Fock space may be orthogonal (as in WT, BT, FBT spaces), but also may not be orthogonal, as in Document Fock space or the TF-space in **LSI**, where we have *applied Dirac notation to SVD*, *derived a **Riemann metric tensor*** and used it to rate documents with respect to a query for the Grossman-Frieder example. We also proposed to use the same metric to define the distances between documents, mapped to a unit sphere in the metric space. We hope our discussion would help to understand, unify and simplify the representations and manipulations of the term, document and query vectors in modern information retrieval.

**Appendix A: The Grossman-Frieder example**

In reference [3], a simple example is used throughout that book. The example, referred to *GF-Example* in this article, has a collection of three Documents and one Query as follows:

$Q$ : "*gold silver truck*"
$D_1$: "Shipment of gold damaged in a fire"
$D_2$: "Delivery of silver arrived in a silver truck"
$D_3$: "Shipment of gold arrived in a truck"

Then the parameters defined in Eq. (1.1.3) can be calculated and listed in Table 1.1.

Table A.1: Term Frequencies ($tf_{i,\mu}$, $tf_{i,q}$ and $idf_i$) and Other Parameters

| Term $i$ | 1 | 2 | 3 | 4 | 5 | 6 | 7 | 8 | 9 | 10 | 11 |
|---|---|---|---|---|---|---|---|---|---|---|---|
| Word | a | arrived | damaged | delivery | fire | **gold** | in | of | shipment | **silver** | **truck** |
| $tf_{i,1}$ | 1 | 0 | 1 | 0 | 1 | 1 | 1 | 1 | 1 | 0 | 0 |
| $tf_{i,2}$ | 1 | 1 | 0 | 1 | 0 | 0 | 1 | 1 | 0 | 2 | 1 |
| $tf_{i,3}$ | 1 | 1 | 0 | 0 | 0 | 1 | 1 | 1 | 1 | 0 | 1 |
| $n_i$ | 3 | 2 | 1 | 1 | 1 | 2 | 3 | 3 | 2 | 1 | 2 |
| $idf_i$ | 0 | 0.176 | 0.477 | 0.477 | 0.477 | 0.176 | 0 | 0 | 0.176 | 0.477 | 0.176 |
| $Q$ | 0 | 0 | 0 | 0 | 0 | 1 | 0 | 0 | 0 | 1 | 1 |
| $tf_{i,q}$ | 0 | 0 | 0 | 0 | 0 | 1 | 0 | 0 | 0 | 1 | 1 |

Here, we calculated **idf** using equation (1.1.3a) with $N = 3$. Readers can get more detailed information in Ref. [3] about calculating weights for this example using Eq. (1.3.3) ([3], pages 16).

[2]. Ricardo Baeza-Yates and Berthier Riberto-Neto. ***Modern Information Retrieval***. Addison Wesley, New York, 1999.

[3]. David A. Grossman and Ophir Frieder. ***Information Retrieval***, *Algorithm and Heuristics*, 2$^{nd}$ Edition. Springer, 2004.

[4]. G. Salton and M.E. Lesk, *Computer evaluation of indexing and text processing*. Journal of the ACM, 15 (1): 8-16, Jan., 1968.

[5]. G. Dalton. *The SMART Retrieval System – Experiments in Automatic Document Processing*. Prentice Hall Inc., Englewood Cliffs, NJ, 1971.

[6]. S. K. M. Wong, W. Ziarko, and P. C. Wong, *Generalized vector space model in information retrieval*. In *Proc. 8$^{th}$ ACM SIGIR Conference on Research and Development in Information Retrieval*, pages 18-25, New York, USA, 1985.

[7]. Gerard Salton, Edward A. Fox, and Harry Wu, *Extended Boolean information retrieval*. Communications of the ACM, 26(11):1022-1036, November 1983.

[8]. Gerard Salton. ***Automatic Text Processing***. Addison-Wesley, 1989.

[9]. Stephen Willard. ***General Topology***, page 16, Dover Publications Inc., New York, 2004. Also see: http://en.wikipedia.org/wiki/Distance#Norms

[10]. Yonggang Qiu and H. P. Frei. *Concept based query expansion*. In *Proc. 16$^{th}$ ACM SIGIR Conference on Research and Development in Information Retrieval*, pages 160-169, Pittsburg, USA, 1993

[11]. G. W. Furnas, S. Deerwester, S. T. Dumais, T. K. Landauwer, R. A. Harshman, L. A. Streeter, and K. E. Lochbaum. *Information retrieval using a singular value decomposition model of latent semantic structure*. In *Proc. of the 11$^{th}$ annual International ACM SIGIR conference on Research and Development in Information Retrieval*, pages 465-480, 1988.

[12]. E. Gasia. *SVD and LSI Tutorial 1-4*. Starting from: http://www.miislita.com/information-retrieval-tutorial/svd-lsi-tutorial-1-understanding.html

[13]. Steven J. Leon. ***Linear Algebra with Applications***. 4$^{th}$ edition, Macmillan College Publishing Company, Inc, 1994.

[14]. Matrix multiplication: http://www.uni-bonn.de/~manfear/matrixcalc.php

[15]. Matrix eigenvalues and eigenvectors: http://www.bluebit.gr/matrix-calculator/

[16]. Riemann Metric: http://en.wikipedia.org/wiki/Riemannian_metric

[17]. S. Weinberg. ***Gravitation and Cosmology***. John Wiley, 1972.

[18]. A Java program: http://www.shermanlab.com/science/CS/IR/PlayMatrix.java

[19]. Keith Van Rijsbergen. ***The Geometry of Information Retrieval***. Cambridge, 2004.

[20]. Lorenzo Sadun. ***Applied Linear Algebra: The Decoupling Principle.*** Prentice Hall, 2000.

[21]. S. Lafon et al., *Diffusion Maps and Coarse-Graining: A Unified Framework for dimensionality Reduction, Graph Partitioning and Data Set Parameterization*. Pattern Analysis and Machine Intelligence, IEEE Transactions on Volume 28, Issue 9, Sept. 2006 Page(s): 1393 - 1403

Dr. Xing M Wang    Dirac, Fock, Riemann and IR    Page 31 of 31